\documentclass[prb,aps, babel,twocolumn]{revtex4}
\usepackage{amsmath,amssymb,amstext,amscd,amsthm,amsopn,graphicx,indentfirst,mathrsfs}

\graphicspath{{.}{./EPS/}}

\begin{document}

\title{Quantum structural phase transition in chains of interacting atoms}

\author{Efrat Shimshoni,$^1$ Giovanna Morigi,$^{2,3}$ and Shmuel Fishman$^4$}

\affiliation{
$^1$ Department of Physics, Bar-Ilan University, Ramat-Gan 52900, Israel\\
$^2$ Theoretische Physik, Universit{\"a}t des Saarlandes, D 66041 Saarbr{\"u}cken, Germany\\
$^3$ Department de F\'isica, Universitat Aut\`onoma de Barcelona, E 08193 Bellaterra, Spain\\
$^4$ Department of Physics, Technion, Haifa 32000, Israel
}
\date{\today}

\begin{abstract}
A quasi one--dimensional system of trapped,  repulsively
interacting atoms (e.g., an ion chain) exhibits a structural phase
transition from a linear chain to a zigzag structure, tuned by
reducing the transverse trap potential or increasing the particle
density. Since it is a one dimensional transition, it takes place
at zero temperature and therefore quantum fluctuations dominate.
In [Fishman, {\it et al.}, Phys. Rev. B  {\bf 77}, 064111 (2008)] it was shown that the system
close to the linear-zigzag instability is described by a $\phi^4$ model.
We propose a mapping of the $\phi^4$ field theory to the well
known Ising chain in a transverse field, which exhibits a quantum
critical point. Based on this mapping, we estimate the quantum critical point in
terms of the system parameters. This estimate gives the critical
value of the transverse trap frequency for which the quantum phase
transition occurs, and which has a finite, measurable deviation
from the  critical point evaluated within the classical theory. A
measurement is suggested for atomic systems which can probe the critical trap
frequency at sufficiently low temperatures $T$. We focus in particular on
a trapped ion system, and estimate the implied limitations on $T$ and on the interparticle distance.
We conclude that the experimental observation of the quantum critical
behavior is in principle accessible.
\end{abstract}


\maketitle

\section{Introduction}
\label{sec:intro}

The structural transition from a linear chain of  repulsively
interacting particles to a planar configuration, in the form of a
zigzag structure, has been often discussed in theoretical studies
on atomic and condensed matter systems. Examples include electrons
in nanowires~\cite{MML}, ultracold dipolar gases~\cite{Kollath},
vortex lines in Bose-Einstein condensates~\cite{Santos,Busch}, and ion
Coulomb crystals in traps~\cite{Hasse,Dubin93,Schiffer93}.
Specifically in ion Coulomb crystals this transition has been
experimentally observed and
characterized~\cite{Dubin,Walther,Wineland}, thereby determining a
phase diagram of the ionic structures as a function of the trap
aspect ratio and of the mean interparticle distance~\cite{Walther}.

Theoretical studies demonstrated that the transition from the
string to the zigzag chain is associated with a symmetry breaking.
 More specifically, if the Hamiltonian is invariant under
rotation about the string axis, it is the rotational symmetry
around the chain which is broken in the zigzag phase. If instead
the motion of the interacting particles is confined to the
plane, the zigzag phase breaks the symmetry by reflection about
the chain axis.  The phase transition predicted by the classical
theory is hence second-order~\cite{Piacente,FCCM}. It can be
described by a Landau model, of which critical values and
exponents are well known~\cite{FCCM}, and where the soft mode is
the zigzag mode of the linear chain -- namely, the transverse mode
with the shorthest wavelength -- which drives the instability and
determines the new structure. The corresponding Ginzburg-Landau
equation in the continuum limit and in presence of damping was
reported in Ref.~\onlinecite{DelCampo}, where special focus was given on
creation of defects when quenching the value of the transverse
trap frequency across the mechanical instability. Further works
discussed the structural transitions when the external potential
is not harmonic~\cite{Peeters09}, showing that the statistical
mechanics of the system at the instability may be profoundly
modified. The results of these studies are strictly valid in the
classical regime, when the thermal fluctuations can be neglected.

\begin{figure}
\centerline{
\includegraphics[angle=270,width=0.6\linewidth]{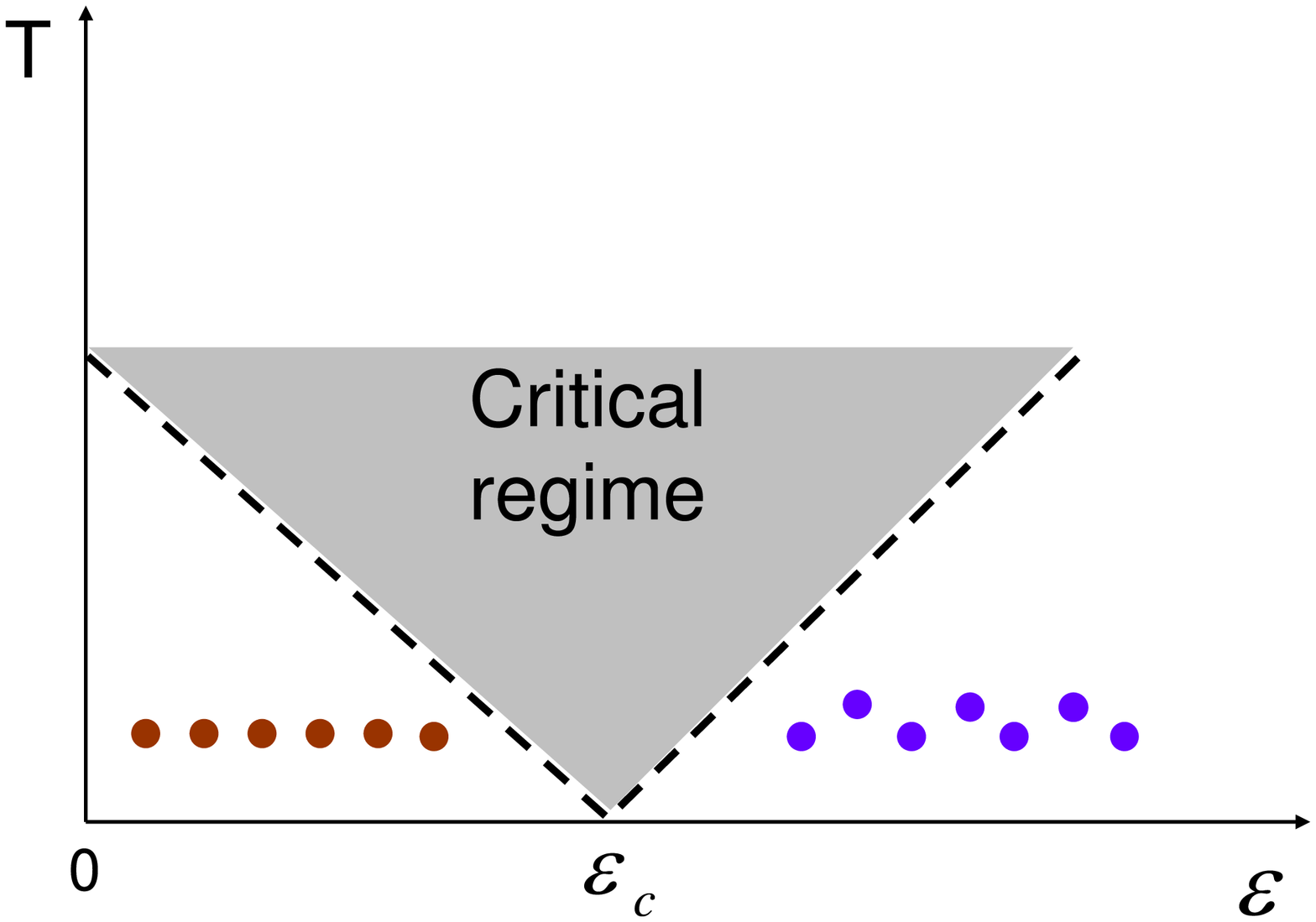}}
\caption{(color online) Sketch of the phase diagram exhibiting a linear-zigzag transition, according to the mapping to the 1D model of a quantum Ising transition. Here, $T$ is the temperature of the sample and the dimensionless parameter $\epsilon$ is tuned by the confining potential or the interparticle distance. The value $\epsilon=0$ corresponds to the value of the transverse frequency, where the linear chain is classically unstable~\cite{FCCM}. The quantum critical point, at $\epsilon_c>0$, separates the linear chain from the zigzag phase at $T=0$. For $0<\epsilon<\epsilon_c$ quantum fluctuations dominate, and the crystal is in the linear (disordered) phase. The dashed lines indicate the boundaries of the quantum critical region, where thermal fluctuations dominate.} \label{Fig:1}
\end{figure}

In Ref.~\onlinecite{SMF} we addressed the question, whether quantum
fluctuations significantly modify the location of the critical
point and the behavior in its vicinity, focussing on ion Coulomb crystals.
This question draws on
numerical studies in low dimensional dipolar systems, which showed
a significant contribution of quantum fluctuations at the
transition from a string to a zigzag order~\cite{Astrakharchik}.
In this article we provide the detailed calculations at the basis of the results presented in Ref.~\onlinecite{SMF},
and extend the treatment to interacting atomic chains, with repulsive interaction potential scaling with $1/x^\alpha$ with $\alpha\ge 1$.
We focus on the case where the motion of the particles is
confined to a plane. As argued below, the string-to-zigzag
transition is indeed a quantum phase transition analogous to the
ferromagnetic transition of an Ising chain in a transverse
field~\cite{QPTbook}. A phase diagram summarizing the behavior in
its vicinity is depicted in Fig.~\ref{Fig:1}. Note that a
similar mapping to the Ising model was argued in the context of
electronic systems in quantum wires~\cite{MML}. However, the
validity of the assumed equivalence between a continuous field
theoretical model and a discrete ($Z_2$) field-theory cannot be
justified on the basis of classification into universality classes
below three space dimensions~\cite{Wilson}.

In the present study we develop a quantum field
theoretical approach, which allows a direct mapping to the quantum one--dimensional Ising model
under plausible assumptions. On this basis we determine the quantum critical point at which the transition occurs and identify the experimental parameters, for which it can be distinguished from the value predicted by the classical theory~\cite{FCCM}. In particular, we relate the critical value of the transverse trap frequency $\nu_t$ to the ratio of the typical kinetic energy scale
$U_K$ and interaction energy $U_P$. Defining a parameter $\epsilon=1-\nu_t^2/\nu_c^2$, which describes the
deviation of the transverse confinement from its stability point, we find that the critical point is given by $\epsilon_c\sim (U_K/U_P)^{2/3}$. A measurement is suggested, which can demonstrate transverse quantum fluctuations in a long chain of trapped ions (composed of hundreds of ions) and probe the different regimes in the phase diagram.

This paper is organized as follows. In Sec. \ref{sec:model} we derive a quantum field theory for the zigzag phonon mode. The mapping to the quantum Ising model is described in Sec. \ref{sec:mapping}. In Sec. \ref{sec:QCP} we derive expressions for the Ising model parameters, and in particular their critical values, in terms of microscopic parameters of the interacting atoms system. In Sec. \ref{sec:experiment}
we discuss  an experimental realization in trapped ions systems, and specify the conditions under which the quantum phase transition is likely to be observable. Finally, our conclusions and outlook are summarized in Sec. \ref{sec:conclude}.

\section{Quantum Field Theory for the Zigzag Mode}
\label{sec:model}

We start by considering a linear array of particles with mass $m$,
which are confined in the $x-y$ plane. The system is at low
temperature, and the particles at equilibrium are aligned along
the $x$-axis and equidistant, with interparticle distance $a$ such
that their equilibrium position is $x_j^{(0)}=ja$ and
$y_j^{(0)}=0$, while the motion along the $z$-axis is assumed to
be frozen out. Periodic boundary conditions are assumed at the
chain edges and one ion of the chain is assumed to be pinned. The
particles can oscillate around the equilibrium positions in the
$x$ and $y$ direction, where they are confined by a harmonic
potential of the form
\begin{equation}
V_t=\frac{1}{2}m\nu_t^2\sum_{j=1}^N y_j^2\; ;
\label{Vt}
\end{equation}
here $y_j$ is the displacement of the $j$'th atom from the chain
axis and $\nu_t$ the trap frequency. Finally, they interact by
means of the repulsive power law potential,
\begin{equation}
V_{int}=\frac{A^2}{2}\sum_{i\not=j}\frac{1}{[(x_i-x_j)^2+(y_i-y_j)^2]^{\alpha/2}}
\label{Vint}
\end{equation}
where $\{{\bf r}_j=(x_j,y_j)\}$ denotes the atomic position in the plane, $A^2$ is the strength of the interaction and $\alpha\ge 1$.

This model includes the Coulomb interaction when $\alpha=1$ and the dipolar interaction when $\alpha=3$, assuming that the dipoles are polarized along the $z$ axis, such that in the $x-y$ plane they experience an isotropic repulsive force. We also note that in the case of the Coulomb interaction the strength is such that $A=Q$, where $Q$ is the charge of the particle. In this case the array can result from the self-organization of the atoms, since quasi-long-range order can be assumed as long as one particle in the chain is pinned. The configuration, in which the particles are aligned along the $x$-axis with uniform interparticle distance, could be realized in a ring trap at ultralow temperatures  under the condition that one ion is  pinned~\cite{Walther}, or in a anharmonic linear Paul trap~\cite{Monroe}. It also corresponds with good approximation to the ions distribution at the center of a long chain in a linear Paul trap~\cite{Walther,Drewsen,FCCM,MF}.

For $\alpha>1$, on the other hand, particle-particle correlations
decay at $T=0$ with a power law dependence on the distance~\cite{Giamarchi}, and long-range
order may be only assumed in presence of an additional periodic
potential along $x$, which localizes the particles at the
positions $x_j^{(0)}$. For dipolar atoms or molecules with an
electric dipole moment $D$, this configuration could be
realized by means of an optical lattice~\cite{Lahaye} with
periodicity $a$. Here, the dipole-dipole interaction can be made repulsive in the $x-y$ plane
when a static electric field along the $z$ axis aligns the dipole. Then, the strength of the
interaction is such that $A\propto D$.

We assume temperatures such that the atoms crystallize and their
motion is well approximated by harmonic vibrations about the
equilibrium positions when the configuration is mechanically
stable. While the linear chain is stable against longitudinal
displacements from the equilibrium positions, a soft mode in the
transverse direction drives a structural phase transition of the
linear chain into a zigzag configuration, as shown in
Ref.~\onlinecite{FCCM}. This mode has wavelength $\lambda=2a$, such that
in the classical harmonic crystal the corresponding displacements
are described by the function $y_j=(-1)^jy_0$ with $y_0$ the
amplitude of the oscillations. We therefore focus on the dynamics
of the transverse phonon modes close to the instability and expand
$V_{int}$ to fourth order in $|y_j|\ll a$.  Moreover, we introduce
the field $\phi_j$, which describes a transverse excitation close
to the instability and is slowly varying with respect to the
length scale $a$, i.e.
\begin{equation}
\phi_j=(-)^jy_j/a\,.
\end{equation}
Using a gradient expansion to leading order~\cite{FCCM}, we obtain
the effective potential
\begin{widetext}
\begin{eqnarray}
V[\{\phi_j\}]=V_t+V_{int}
\approx \frac{1}{2}m\nu_t^2a^2\sum_{j=1}^N \phi_j^2
 + \frac{A^2}{2a^\alpha}\sum_{i\not=j}\frac{\alpha}{2}\left\{-\frac{(\phi_i-(-1)^{i-j}\phi_j)^2}{|i-j|^{\alpha+2}}
+\frac{(\alpha+2)(\phi_i-(-1)^{i-j}\phi_j)^4}{4|i-j|^{\alpha+4}}\right\}\label{V_t_tot} \; .
\end{eqnarray}
\end{widetext}
Terms describing the coupling between the axial and transverse
modes have not been reported, because they give rise to higher
order corrections in the gradient expansion. The latter is defined
by the small parameter $\delta k a\ll 1$, with $\delta k=|k-k_0|$
the typical deviation between the wave vector $k$ of the excited
mode and the wave vector $k_0=\pi/a$ of the soft mode. An
extensive discussion on the derivation of the effective potential
for the soft mode can be found in Ref.~\onlinecite{FCCM}.  The
ordered, zigzag phase is established when $\phi_j$ acquires a
finite, constant expectation value, as we show below. To leading
order in a gradient expansion, the potential  in
Eq.~(\ref{V_t_tot}) can be mapped to the dynamics described by a
potential of the form
\begin{equation}
V[\{\phi_j\}]\approx \sum_{j=1}^{N} V_0(\phi_j) + \frac{1}{2}K\sum_{j=1}^N (\phi_j-\phi_{j+1})^2\; ,
\label{V_phi}
\end{equation}
where the interaction is now nearest neighbors.
Here $V_0$ is  a local potential
\begin{equation}
V_0(\phi)=-\frac{1}{2}m(\nu_c^2-\nu_t^2)a^2\phi^2+\frac{1}{4}ga^4\phi^4\; ,
\label{V_loc}
\end{equation}
and the parameters $K$, $\nu_c$ and $g$ are given by
\begin{eqnarray}
K&= &\frac{C_1[\alpha]\alpha A^2}{a^\alpha}\; ,\quad
\nu_c^2=\frac{4C_2[\alpha+2]\alpha A^2}{ma^{\alpha+2}} \; ,\nonumber\\
g&=& \frac{8C_2[\alpha+4]\alpha(\alpha+2) A^2}{a^{\alpha+4}}\; , \label{V_parameters}
\end{eqnarray}
with
\begin{eqnarray}
C_1[\alpha]\equiv \sum_{j=1}^N \frac{(-)^{j+1}}{j^\alpha}\; ,\quad C_2[\alpha]\equiv \sum_{j\geq 1\, ,odd} \frac{1}{j^\alpha}\; .\nonumber
\end{eqnarray}
A classical theory, which neglects quantum fluctuations (i.e.,
valid in the limit $\hbar=0$), predicts a transition from a linear
to a zigzag chain at $T=0$ when the transverse
confining frequency $\nu_t$ is reduced below the critical value
$\nu_c$~\cite{FCCM}. For later convenience we define a
dimensionless parameter describing the deviation  of the
transverse trap frequency from this classical transition point:
\begin{equation}
\epsilon\equiv \frac{\nu_c^2-\nu_t^2}{\nu_c^2}\; .
\label{def_epsilon}
\end{equation}
For $\epsilon >0$, the local potential $V_0$ [Eq. (\ref{V_loc})] has minima at
\begin{equation}
\phi_\pm=\pm \phi_0\; ,\quad
\phi_0\equiv\frac{1}{a}\sqrt{\frac{m\omega^2}{g}}\; ,
\label{phi_0}
\end{equation}
where the frequency $\omega$ is given by the relation
\begin{equation}
\omega^2=\nu_c^2-\nu_t^2=\epsilon\nu_c^2
\label{omega_def}
\end{equation}
and is associated with the curvature of $V_0$ at the top of the barrier between the minima of the double well potential, as one can observe from Eq.~(\ref{V_loc}).

We note that the mapping to Eq.~(\ref{V_phi}) is valid for a
power-law interaction with $\alpha\ge 1$ (for which the relevant
sums $C_l[\alpha]$ in Eq.~(\ref{V_parameters}) converge), hence
including the case of the Coulomb interaction at $\alpha=1$.
Therefore, for the linear-zigzag instability the Coulomb
interaction effectively belongs to the class of short-range
potentials. Indeed, one finds that the spectrum of the transverse
excitations close to the instability (approaching from the side
where the chain is stable classically) behaves as $$\omega^2 \sim
\nu_t^2-\nu_c^2+(K/m)k^2$$ for $\alpha\ge 1$ (see for instance
Ref.~\onlinecite{FCCM} for the case $\alpha=1$). This behavior is
markedly different from that of the long wavelength, axial
excitations: for $\alpha>1$ the axial frequency $\omega\sim ck$,
with $c$ sound velocity, while for $\alpha=1$ there exists no
sound velocity, and $\omega \sim k\sqrt{|\log
k|}$~\cite{Ashcroft,MF,MF06}.

These considerations on the structure, which are based on identifying the minima of
the potential energy, do not account for the contribution of the
kinetic energy. Thermal effects are expected to modify the
behaviour at the critical point. Moreover, even at temperatures
$T\rightarrow 0$, when thermal fluctuations are small, quantum
fluctuations will become relevant. Sufficiently close to the value
$\nu_c$, i.e. for sufficiently small values of the parameter $\epsilon$ (where the classical zigzag ordering is expected), quantum
fluctuations will induce tunneling between degenerate minima of the potential and
are expected to destroy the zigzag ordering. A true phase
transition will therefore occur at a smaller value of $\nu_t$,
corresponding to a {\it quantum} critical point, such that
$\nu_t<\nu_c$; i.e., at $\epsilon_c >0$.

In order to investigate the quantum critical behavior, one has to introduce a 1+1 dimensional quantum field--theory for the system. To this end, we write the partition function as
\begin{equation}
Z=\int {\mathcal D}\phi\, e^{-S[\phi]/\hbar}
\label{Z_def}
\end{equation}
with the Euclidean action
\begin{eqnarray}
S[\phi]&=& \int_0^{\hbar\beta} d\tau \sum_{j=1}^N
\left[\frac{1}{2}ma^2(\partial_\tau \phi_j)^2\right. \nonumber\\
&+&V_0(\phi_j)+\left.\frac{1}{2}K(\phi_j-\phi_{j+1})^2\right]\; ,
\label{S_phonons}
\end{eqnarray}
where $\beta=1/k_BT$ and $\beta\rightarrow \infty$ for $T\rightarrow 0$. Below we demonstrate the mapping of this model to the Ising chain in a transverse field, and derive approximate expressions for its parameters in terms of microscopic parameters of the interacting particles system. This will facilitate the study of the conditions under which the phase transition can be observable in such systems.

\section{Mapping to the Quantum Ising Chain}
\label{sec:mapping}

The low--energy model derived in the previous section [Eqs. (\ref{Z_def}), (\ref{S_phonons})] describes the quantum dynamics of the zigzag phonon mode in terms of a real continuous scalar field $-\infty<\phi_j(\tau)<\infty$. We now express it in a form which will allow its mapping onto an effective model for the discrete field $\sigma_j(\tau)\equiv {\rm Sgn}[\phi_j(\tau)]$. As a starting point, we perform the standard procedure of dividing the imaginary time ($\tau$) axis into discrete time steps $\{\tau_m\}$ separated by an infinitesimal interval of size $\delta\tau$, with $m=0,\ldots,M$ and $M=\hbar\beta/\delta\tau$. The partition function is then cast in the form
\begin{equation}
Z={\rm Tr}\,\left\{\hat T^M\right\}
\label{Z_T}
\end{equation}
where $\hat T$ is a transfer matrix describing the time evolution from $\tau_m$ to $\tau_{m+1}$. Using the notation $\phi_j=\phi_j(\tau_m)$, $\phi^\prime_j=\phi_j(\tau_{m+1})$, the matrix elements of $\hat T$ are given by
\begin{widetext}
\begin{eqnarray}
\label{T_mat}
T[\{\phi_j\},\{\phi^\prime_j\}]&=&\exp\left\{-\frac{\delta\tau}{2\hbar}
\sum_{j=1}^N\left[ma^2\left(\frac{\phi_j-\phi^\prime_j}{\delta\tau}\right)^2 + V_0(\phi_j)+V_0(\phi^\prime_j)
+\frac{1}{2}K\left((\phi_j-\phi_{j+1})^2+(\phi^\prime_j-\phi^\prime_{j+1})^2\right)\right]\right\}\;\\
&=&\prod_{j=1}^NG(\phi_j,\phi^\prime_j;\delta\tau)\;\exp\left\{-\frac{\delta\tau}{4\hbar}K
\sum_{j=1}^N\left[(\phi_j-\phi_{j+1})^2+(\phi^\prime_j-\phi^\prime_{j+1})^2\right]\right\}\; .
\label{T_loc}
\end{eqnarray}
\end{widetext}
Here the propagator $G(\phi_j,\phi^\prime_j;\delta\tau)$ is given by
\begin{equation}
G(\phi_j,\phi^\prime_j;\delta\tau)=\langle \phi^\prime_j
|e^{-{\hat H_j}\delta\tau/\hbar}|\phi_j\rangle\; ,
\label{prop_def}
\end{equation}
in which the local Hamiltonian $\hat H_j=p_j^2/(2m)+V_0$ describes
the quantum dynamics of a particle in the double--well potential
$V_0(\phi)$, where $p_j$ is the momentum conjugate to
$\phi_j$. The symmetry of the potential implies that the exact
eigenstates of $\hat H_j$ consist of pairs of states with
well-defined symmetry under $\phi\rightarrow -\phi$: symmetric
($|S_n\rangle$) and antisymmetric ($|A_n\rangle$). The
corresponding energy eigenvalues can be expressed as $\hbar({\bar
\omega}_n-\Delta \omega_n/2)$, $\hbar({\bar \omega}_n+\Delta
\omega_n/2)$ respectively, where the splitting energy $\hbar\Delta
\omega_n$ is associated with the overlap of wave functions
centered in either of the two wells. The propagator
$G(\phi,\phi^\prime;\delta\tau)$ (the index $j$ is dropped to
simplify the notation) therefore acquires the form
\begin{eqnarray}
& &G(\phi,\phi^\prime;\delta\tau) =\sum_n e^{-{\bar
\omega}_n\delta\tau} \left\{\langle \phi^\prime |S_n\rangle
\langle S_n|\phi\rangle e^{\Delta
\omega_n\delta\tau/2}\right.\nonumber\\&+&\langle \phi^\prime
|A_n\rangle \langle A_n|\phi\rangle \left. e^{-\Delta
\omega_n\delta\tau/2}\right\}\; . \label{prop_SA}
\end{eqnarray}

We now change basis into wave -functions centered at the right and
left potential minima $\phi_\pm$ [Eq. (\ref{phi_0})]:
\begin{equation}
|R_n\rangle \equiv \frac{1}{\sqrt{2}}(|S_n\rangle + |A_n\rangle )\; ,\quad
|L_n\rangle \equiv \frac{1}{\sqrt{2}}(|S_n\rangle - |A_n\rangle )\; .
\label{RL_def}
\end{equation}
Equation~(\ref{prop_SA}) becomes
\begin{eqnarray}
& &G(\phi,\phi^\prime;\delta\tau) =\sum_n e^{-{\bar
\omega}_n\delta\tau} \label{prop_RL}\\ &\times
&\left\{\left[\langle \phi^\prime |R_n\rangle \langle
R_n|\phi\rangle + \langle \phi^\prime |L_n\rangle \langle
L_n|\phi\rangle \right] \cosh\left\{\frac{\Delta
\omega_n\delta\tau}{2}\right\}\right.\nonumber\\&+&\left[\langle
\phi^\prime |R_n\rangle \langle L_n|\phi\rangle +\langle
\phi^\prime |L_n\rangle \langle R_n|\phi\rangle\right]\left.
\sinh\left\{\frac{\Delta \omega_n\delta\tau}{2}\right\}\right\}\;
. \nonumber
\end{eqnarray}
We next define the wave functions $\Psi_{n,\zeta}(\phi)\equiv \langle \phi |\zeta_n\rangle$, where $\zeta =+,-$ denotes the isospin $R,L$, respectively. Since these wave functions are not known exactly, we implement a variational approach and assume the trial function
\begin{equation}
\Psi_{n,\zeta}(\phi)=f_n(\phi)\exp\left\{-\frac{(\phi-\zeta\phi_0)^2}{2l_n^2} \right\}\; ,
\label{RL_wf}
\end{equation}
where $f_n(\phi)$ for $n\geq 1$ is an oscillatory, symmetric
function of $\phi$ with a number of nodes increasing with $n$; for
the lowest energy states $n=0$ we assume
$f_0(\phi)=1/(\sqrt{\pi}l)^{1/2}$, so that $l\equiv l_0$ is a
single variational parameter. The dependence of $\Psi_{n,\zeta}(\phi)$ on the magnitude and sign of $\phi$ can be made explicit
using the substitution $\phi =|\phi|\sigma$, which yields
\begin{eqnarray}
\Psi_{n,\zeta}(\phi)&=&f_n(|\phi|)\exp\left\{-\frac{(|\phi|-\phi_0)^2}{2l_n^2}\right\} \label{wf_sigma}\\ &\times &\left(\delta_{\sigma\zeta}+\exp\left\{\frac{-2|\phi|\phi_0}{l_n^2}\right\}\delta_{\sigma,-\zeta}\right)\; . \nonumber
\end{eqnarray}
The first term in Eq. (\ref{wf_sigma}) dominates as long as $\phi_0>l_n$.

To complete the derivation of an effective field theory in terms
of the discrete fields $\sigma_j(\tau)$, one needs to perform the
path integral [Eq. (\ref{Z_def})] over the magnitude-field
$|\phi_j(\tau)|$. Due to the Gaussian factor in Eq.
(\ref{wf_sigma}), the integration over $\phi_j\equiv\phi_j(\tau)$
(for each $n$) can be written schematically as
\begin{equation}
\int d\phi_j\exp\left\{-\frac{(\phi_j-\sigma_j\phi_0)^2}{2l_n^2}\right\}\exp\{F(\phi_j)\delta\tau\}
\label{int_phi}
\end{equation}
where $F(\phi_j)$ encodes the remaining $\phi_j$-dependence, including in particular the interparticle coupling terms $(K/2)(\phi_j-\phi_{j\pm 1})^2$ [the last term in Eq. (\ref{S_phonons})].
If we now consider the case where $\phi_0\gg l_n$ (which is valid deep in the zigzag phase), a saddle-point approximation of the integral yields (up to a multiplicative constant factor) $e^{F(\phi_0\sigma_j)\delta\tau}$. This implies that the partition function [Eq. (\ref{Z_T})] can be recast as
\begin{equation}
Z={\rm Tr}\left\{\hat T_{eff}^M\right\}\,,
\label{Z_Teff}
\end{equation}
with ${\rm Tr}=\sum_{\{\sigma_j\}}$ and $\hat T_{eff}\,$ a $\,2^N\times 2^N$--matrix related to the original transfer matrix $\hat T$ by
\begin{equation}
 T_{eff}[\{\sigma_j\},\{\sigma^\prime_j\}]=T[\{\phi_0\sigma_j\},\{\phi_0\sigma^\prime_j\}]\; .
\label{Teff_def}
\end{equation}
Recalling Eq. (\ref{T_mat}), we find that $T_{eff}[\{\sigma_j\},\{\sigma^\prime_j\}]$ can be expressed as
\begin{eqnarray}
& & T_{eff}[\{\sigma_j\},\{\sigma^\prime_j\}] \label{Teff}\\&=&\prod_{j=1}^N T_{loc}[\sigma_j,\sigma^\prime_j]\exp\left\{\frac{K\phi_0^2\delta\tau}{2\hbar}\left(\sigma_j\sigma_{j+1}+\sigma^\prime_j\sigma^\prime_{j+1}\right) \right\}\nonumber \;
\end{eqnarray}
where $T_{loc}[\sigma,\sigma^\prime]=const\times G(\phi_0\sigma,\phi_0\sigma^\prime;\delta\tau)$,
and the propagator $G(\phi_0\sigma,\phi_0\sigma^\prime;\delta\tau)$ can be obtained by inserting Eq. (\ref{wf_sigma}) in Eq. (\ref{prop_RL}).

The final stage in the mapping to the quantum Ising chain involves a derivation of an explicit expression for the $\,2\times 2$--matrix ${\hat T}_{loc}$, which encodes the single particle dynamics. By inspection of Eqs. (\ref{wf_sigma}) and (\ref{prop_RL}) it is evident that ${\hat T}_{loc}$ obeys the symmetry $T_{loc}[\sigma,\sigma^\prime]=T_{loc}[-\sigma,-\sigma^\prime]$. It can therefore be written in the form
\begin{equation}
{\hat T}_{loc}=const\times (A_0\sigma^0+A_X\sigma^x)\; ,
\label{Tloc_gen}
\end{equation}
where $\sigma^\alpha$ are Pauli matrices in the basis where
$\{\sigma=\pm 1\}$ denote the eigenvalues of $\sigma^z$, and
$\sigma^0$ is the $\,2\times 2$ unit matrix. This is the
crucial point where we formulate the symmetry of the problem,
which stems from the symmetry relations leading to Eqs.
(\ref{prop_def})-(\ref{RL_def}). The general expressions for
$A_0$ and $A_X$ are
\begin{widetext}
\begin{eqnarray}
A_0&=&\sum_n e^{-{\bar\omega}_n\delta\tau}
f_n^2(\phi_0)\left(\cosh\left\{\frac{\Delta
\omega_n\delta\tau}{2}\right\}+\Gamma^2 \sinh\left\{\frac{\Delta
\omega_n\delta\tau}{2}\right\}\right)\label{AandB}\\
A_X&=&\sum_n e^{-{\bar\omega}_n\delta\tau}
f_n^2(\phi_0)\left(\sinh\left\{\frac{\Delta
\omega_n\delta\tau}{2}\right\}+\Gamma^2 \cosh\left\{\frac{\Delta
\omega_n\delta\tau}{2}\right\}\right)\; , \nonumber
\end{eqnarray}
\end{widetext}
where
\begin{equation}
\Gamma\equiv e^{-\phi_0^2/l_n^2}\; . \label{overlap}
\end{equation}

Equation (\ref{AandB}) implies that in general $A_0$ and $A_X$ are
non--trivial functions of the time interval $\delta\tau$.
Restrictions on the parameters must therefore be set in order to
establish a well--defined effective model for arbitrarily small
$\delta\tau$. A simplification can be achieved in view of the fact
that the sum over $n$ is rapidly converging: a semiclassical
calculation indicates that $\omega_n\sim n^{4/3}$ for large $n$
(see Appendix for details), i.e. the level spacing increases with $n$.
In the regime where the tunnel splitting obeys $\Delta\omega_n\ll
{\bar\omega}_n$ (realized far enough from the
 quantum transition so that the wells of $V_0$ are well separated), the
sum is dominated by the first term $n=0$ up to exponentially small
corrections. Even closer to the phase transition, where the
tunnel splitting becomes comparable to the barrier between the
wells, the sum in Eq. (\ref{AandB}) is dominated by the first few
terms. As we show in the next section, provided the quartic term
in $V_0$ (i.e., the parameter $g$) is sufficiently large, all the
energy levels $\hbar({\bar\omega}_n\pm \Delta\omega_n)$ are given
(up to a numerical factor) by {\it the same} energy scale
$(\hbar^4g/m^2)^{1/3}$. Since the differences
$(\Delta\omega_n-\Delta\omega_{n-1})\sim n^{-2/3}$ are typically
smaller than ${\Delta\omega}_n$, the dependence on $n$ of all
terms in the brackets in Eq. (\ref{AandB}) may be neglected. One
can therefore approximate $\Delta\omega_n$ for all $n$ by
$\Delta\omega\sim (\hbar g/m^2)^{1/3}$. This yields
\begin{widetext}
\begin{equation}
T_{loc}\approx const\times \left[\left(\cosh\left\{\frac{\Delta
\omega\delta\tau}{2}\right\}+\Gamma^2 \sinh\left\{\frac{\Delta
\omega\delta\tau}{2}\right\}\right)\sigma^0+\left(\sinh\left\{\frac{\Delta
\omega\delta\tau}{2}\right\}+\Gamma^2 \cosh\left\{\frac{\Delta
\omega\delta\tau}{2}\right\}\right)\sigma^x\right]\; .
\label{Tloc_n0}
\end{equation}
\end{widetext}
Assuming in addition that the overlap factor $\Gamma$ [Eq. (\ref{overlap})] obeys
\begin{equation}
\Gamma\ll\tanh\{\Delta\omega\delta\tau /2\}\ll 1\; ,
\label{condition}
\end{equation}
Eq. (\ref{Tloc_n0}) reduces to
\begin{eqnarray}
& &T_{loc}[\sigma,\sigma^\prime] \label{Tloc}\\ &\approx &T_0\left(\cosh\left\{\frac{\Delta\omega\delta\tau}{2}\right\}\delta_{\sigma\sigma^\prime}
+\sinh\left\{\frac{\Delta\omega\delta\tau}{2}\right\}\delta_{\sigma,-\sigma^\prime}\right) \nonumber
\end{eqnarray}
where $T_0$ is a constant prefactor. In terms of Pauli's matrices this becomes
\begin{eqnarray}
{\hat T}_{loc}&\approx & T_0\left(\cosh\left\{\frac{\Delta\omega\delta\tau}{2}\right\}\sigma^0
+\sinh\left\{\frac{\Delta\omega\delta\tau}{2}\right\}\sigma^x\right)\nonumber \\
&=& T_0\exp\left\{\frac{\Delta\omega\sigma^x\delta\tau}{2}\right\}\; .
\label{Tloc_sigma_x}
\end{eqnarray}
As a result, the expression for $Z$ [Eqs. (\ref{Z_Teff}) and (\ref{Teff})] effectively reduces to
\begin{equation}
Z\approx Z_0\,\int {\mathcal D}\sigma e^{-S_I[\sigma]/\hbar}
\label{Z_Ising}
\end{equation}
in which $Z_0$ is a constant and $S_I$ is the action of an Ising
chain in a transverse field \cite{QPTbook,GNTbook} subject to the
Hamiltonian
\begin{equation}
H_I=-\sum_{j=1}^N (J\sigma^z_j\sigma^z_{j+1}+h\sigma^x_j)\; .
\label{H_Ising}
\end{equation}
Here, the fictitious exchange coupling $J$ and transverse field $h$ are given by
\begin{equation}
J=K\phi_0^2=\frac{Km\omega^2}{a^2g}\; ,\quad h=\frac{\hbar\Delta\omega}{2}
\label{Ising_Jh}
\end{equation}
where we have used Eq. (\ref{phi_0}) for $\phi_0$. The model described by Hamiltonian~(\ref{H_Ising}) is known to exhibit a quantum phase transition at $h/J=1$ and $T=0$, separating an ordered phase at $J>h$ (in our case, the stable zigzag configuration) from a disordered phase at $J<h$ (the linear chain). In both phases the spectrum of excitations is characterized by a gap with energy $\Delta=2|J-h|$. The corresponding phase diagram,
showing the critical behavior as a function of $\epsilon$ (and hence $\nu_t$) and $T$ is sketched in Fig.~\ref{Fig:1}. The critical regime
($k_B T>\Delta$) is characterized by universal power-law $T$-dependence of correlation functions with the critical exponents of the quantum transition~\cite{QPTbook}. In the next section we evaluate the gap in terms of the atom chain system parameters, and in particular relate it to the parameter $\epsilon$ [Eq.
(\ref{def_epsilon})]. This will enable an estimate of the critical value $\epsilon_c$ satisfying $\Delta(\epsilon_c)=0$.

It should be noted that the condition (\ref{condition}) was
required to obtain the form (\ref{Tloc_sigma_x}) for the local
transfer matrix, and thus the final stage in the derivation of
Eqs. (\ref{Z_Ising}),(\ref{H_Ising}). Namely, the mapping to the
Ising model Eq. (\ref{H_Ising}) is valid
only as long as the lowest energy levels are below the barrier in
the potential $V_0$. In this regime, the tunneling between the
wells is small and hence necessarily $h\ll J$. This is satisfied
deep in the ordered (zigzag) phase, in which case all our
approximations are justified. It does {\it not} hold near the
critical point, which we estimate in the next section. However, we
note that conformal symmetry strongly restricts the number of
universality classes in $1+1$-D ~\cite{Cardy}.  Moreover,
it was argued that the number of relevant operators in two
dimensions is the same as in $4-\epsilon$ dimension~\cite{Cardy},
therefore the resulting $\phi^4$ field theory describing our model
is expected to be in the same universality class as the two dimensional Ising
model. Based on the symmetry properties of the present system
(as manifested by Eq. (\ref{Tloc_gen}) and the following
derivation) it is reasonable to assume that a quantum Ising model
is the appropriate field theory, which correctly describes its
critical behavior at $h\sim J$. This hypothesis is further
supported by numerical studies, e.g., by Barma and
Fisher~\cite{Fisher} and  a later work by Kim, Lin and
Rieger~\cite{Rieger}. Under these assumptions we estimate the
quantum critical point in the regime of parameters where
classically zigzag order is suppressed by quantum tunneling, and
find that it belongs to a universality class~\cite{QPTbook,Cardy}
that differs from the one of the classical Landau model.

\section{Estimate of the Quantum Critical Parameters}
\label{sec:QCP}

As we show below, the effective coupling parameter $h/J$ of the Ising model [Eq. (\ref{Ising_Jh})] can be expressed in terms of two dimensionless parameters characterizing the system: the first is the ratio of energy scales $U_K/U_P$, where
\begin{equation}
U_K\equiv \frac{\hbar^2}{ma^2}
\label{U_K}
\end{equation}
is a typical kinetic energy scale of atoms in the chain, and
\begin{equation}
U_P\equiv \frac{A^2}{a^\alpha}
\label{U_P}
\end{equation}
is an energy scale associated with the interaction. In atomic systems, this ratio can be tuned most effectively by controlling the density of atoms $1/a$, e.g., by means of the lateral confinement. The second dimensionless parameter $\epsilon$ [Eq. (\ref{def_epsilon})] can in principle be
tuned independently by controlling the transverse confining potential frequency $\nu_t$.

The expression for the fictitious exchange coupling $J$ in terms of the above definitions can be obtained directly from Eq. (\ref{Ising_Jh}).
Employing Eqs. (\ref{V_parameters}), (\ref{omega_def}) and (\ref{U_P}), we rewrite the parameters as
\begin{eqnarray}
K&=&C_1[\alpha]\alpha U_P\; ,\quad
m\omega^2=\frac{4C_2[\alpha+2]\alpha U_P\,\epsilon}{a^2} \; ,\nonumber \\
g &=&\frac{8C_2[\alpha+4]\alpha(\alpha+2)U_P}{a^4}\; .
\label{J_dimpar}
\end{eqnarray}
This yields
\begin{equation}
J=C_JU_P\,\epsilon\; \quad {\rm where}\quad C_J\equiv \frac{C_1[\alpha]C_2[\alpha+2]\alpha}{2C_2[\alpha+4](\alpha+2)}\; .
\label{J_UP}
\end{equation}
Note that since we focus on the regime $\epsilon\geq 0$, and by definition $\epsilon \leq 1$, the energy scale $J$ can be tuned continuously from $0$ to $\sim U_P$.
A quantum phase transition is expected to occur at $\epsilon_c$, dictated by the ratio $U_K/U_P$ at which $J$ is equal to the ``transverse field" energy scale $h$.

To get a concrete estimate of $h$ in terms of $U_K$ and $U_P$, we perform a variational calculation of the splitting energy $\hbar\Delta\omega$. To this end, we consider the trial wave functions
\begin{eqnarray}
& &\Psi_{\pm }(\phi)=\frac{1}{(2\sqrt{\pi}l)^{1/2}} \label{pm_wf} \\
&\times &\left(\exp\left\{-\frac{(\phi-\phi_0)^2}{2l^2} \right\}\pm \exp\left\{-\frac{(\phi+\phi_0)^2}{2l^2} \right\}\right)
\nonumber
\end{eqnarray}
for the lowest energy states [i.e., the symmetric and antisymmetric combinations of Eq. (\ref{RL_wf}) with $n=0$], and search for a minimum of the energies
\begin{equation}
E_\pm=\frac{\langle \pm |H_0|\pm \rangle}{\langle \pm |\pm \rangle}
\label{Epm}
\end{equation}
with respect to the variational parameter $l$. Here $H_0$ is the Hamiltonian of a particle of mass $m$ subject to the double--well potential $V_0$, and $\Psi_{\pm }(\phi)=\langle \phi |\pm \rangle$. It is convenient to eliminate the parameters of the quadratic part of $H_0$ by introducing the length scale $\lambda =\sqrt{\hbar/m\omega}$ and the energy scale $\hbar\omega/2$. Using the normalized coordinate $\bar\phi=(a/\lambda)\phi$ we write the Hamiltonian as
\begin{equation}
H_0=\frac{1}{2}\hbar\omega {\bar H}_0\; ,\quad {\bar H}_0=-\frac{d^2}{d\bar\phi^2}-\bar\phi^2+B\bar\phi^4
\label{H0_norm}
\end{equation}
depending on a single dimensionless parameter
\begin{equation}
B\equiv \frac{\hbar g}{2m^2\omega^3}\; .
\label{B_def}
\end{equation}
The nature of the eigenstates and eigenvalues of $H_0$ crucially depends on $B$ being small or large.

The exact minimum condition for $E_+ [l]$ yields a
cumbersome, transcendental equation for the variational parameter
$l$. In particular, it includes the exponential factors
$e^{-\bar{\phi}_0^2/\bar{l}^2}$, with $\bar\phi_0=(a/\lambda)\phi_0$, $\bar{l}=(a/\lambda)l$. These factors are
associated with the overlap of the right and left-centered
wave functions. However, since we are interested in a regime of
parameters close to the critical point (see discussion between
Eqs.~(\ref{h_UKUP}) and (\ref{qcp})), we focus on the approximation of
a ``quasi--flat" potential, corresponding to $B\gg 1$ and
$\bar{\phi}_0^2/\bar{l}^2\ll 1$. Essentially, this regime amounts to a
dominance of the quartic term in the potential, while the barrier
separating the two wells is relatively small. In fact, the parameter $B$ can be written as $B=\hbar\omega/(2\Delta V)$,
where $\Delta V$ is the difference between the energy of the
minima of $V_0$ and the top of the bump separating them. Therefore,
$B$ is the ratio between the zero-point energy
$\hbar\omega/2$ and the potential energy of the barrier.
As becomes clear from the following discussion, the regime $B>1$ is
relevant for the critical point. In this regime also the two
Gaussian functions of Eq.~(\ref{pm_wf}) are expected to mix, therefore
$\bar{\phi}_0^2/\bar{l}^2<1$. To find an analytic expression we take
these inequalities to the extreme. We find from Eq.~(\ref{phi_0})
$\bar{\phi}_0^2=\frac{1}{2B}$. To leading order in $1/B$, this
yields a minimum of $E_+ [l]$ at
\begin{equation}
l_{min}=\frac{\lambda}{a}\bar{l}_{min}\; ,\quad \bar{l}^2_{min}\approx\left(\frac{1}{3B}\right)^{1/3}\; .
\label{l_min}
\end{equation}
Both $E_+ [l]$ and $E_-[l]$ are computed using Eq.~(\ref{Epm}) within
the above approximations, with the value of $l^2_{min}$ which was
found minimizing $E_+ [l]$. We checked that would we extremize
$E_-[l]$, a value with the same dependence on parameters that
differs only by 3\% would be obtained.

 The resulting energy eigenvalues
$E_\pm =E_\pm [l_{min}]$ are separated by a splitting energy
$\Delta E=E_- [l]-E_+ [l]$ where
\begin{equation}
h\approx\frac{\Delta E}{2}=\frac{1}{2}\hbar\omega (3B)^{1/3}\; .
\label{h_B}
\end{equation}
This sets the scale of the transverse field.

Recalling Eq. (\ref{B_def}) for $B$, we note that the dependence
on $\omega$ (and hence on $\epsilon$) cancels out. We thus
get
\begin{eqnarray}
h&\approx &\frac{1}{2}\left(\frac{3\hbar^4 g}{2m^2}\right)^{1/3}=C_h(U_K^2U_P)^{1/3}\; , \nonumber \\
C_h&\equiv &\left(\frac{3C_2[\alpha+4]\alpha(\alpha+2)}{2}\right)^{1/3}
\label{h_UKUP}
\end{eqnarray}
where we have used Eq. (\ref{J_dimpar}) to relate $g$ to $U_P$. It
should be pointed out that (up to a numerical factor close to unity)
the above estimate for $h$ coincides with the energy scale
dictating the higher energy eigenvalues of $H_0$ in the
semiclassical approximation (formally valid for $n\gg 1$):
$E_n\sim n^{4/3}h$ [see Appendix, Eq. (\ref{sen2})]. We hence
conclude that in this regime of parameters, all energy levels are
determined by the same energy scale.

We finally employ  Eqs. (\ref{J_UP}) and (\ref{h_UKUP}) to evaluate the critical value of $\epsilon$ for a given $U_K/U_P$: the condition
\begin{equation}
h=J
\nonumber
\end{equation}
dictates
\begin{equation}
\epsilon_c\approx C_c\left(\frac{U_K}{U_P}\right)^{2/3} ,\quad C_c\equiv \frac{C_h}{C_J}\approx \left(\frac{12(\alpha+2)^4}{\alpha^2}\right)^{1/3}
\label{qcp}
\end{equation}
where we have used the approximation $C_1[\alpha]\sim C_2[\alpha]\sim 1$ for arbitrary $\alpha$. For $\alpha=1$, $\alpha=3$ the numerical prefactor is $C_c\sim 10$. This implies that for given $U_K/U_P\ll 1$, a quantum critical point is typically expected at $\epsilon_c\gg U_K/U_P$.

\section{Experimental realization}
\label{sec:experiment}

\subsection{Ion Coulomb crystals}

An interesting system to study the above predicted quantum
phase transition would be an ion crystal in a quasi
one--dimensional trap
\cite{Walther,Wineland,Cirac,Blatt_Review,Drewsen}. For
singly-ionized alkali-earth metal atoms, the generalized charge
$A$ is the electron charge ($A=e$), and the mass of the ions can
be written as $m=n_Am_p$ where $m_p$ is the proton mass and $n_A$
the atomic number. In these systems it is possible to tune through the critical point by
either controlling the transverse confinement frequency $\nu_t$
(which is typically in the MHz regime), or the spacing between
neighboring ions $a$ which is conveniently measured in units of
micrometers: $a=a_0\times 1\,\mu$m.

To be able to distinguish between the quantum disordered phase and the ordered (zigzag) phase, one must first make sure that the frequency difference $\delta\nu=\nu_c-\nu_t$ implied by our estimated $\epsilon_c$ [via Eq. (\ref{def_epsilon})] is not limited by the experimentally accessible resolution. To this end, it is useful to write $\epsilon_c$ in terms of $n_A$ and $a_0$ defined above. Inserting Eqs. (\ref{U_K}), (\ref{U_P}) and the  numerical values of $\hbar$, $e$ and $m_p$ in Eq. (\ref{qcp}), we obtain
\begin{equation}
\epsilon_c\approx 10^{-4}\frac{1}{(n_Aa_0)^{2/3}}\; .
\label{qcp_num}
\end{equation}
From Eq. (\ref{def_epsilon}), we therefore obtain an upper bound
to the frequency resolution
\begin{equation}
\delta\nu\approx 10^{-4}\frac{1}{2(n_Aa_0)^{2/3}}\nu_c
\label{delta_nu}
\end{equation}
which ranges from 100$\,$Hz (for protons with $a_0\sim 1$) to 1$\,$Hz for heavier ions. This bound has to be compared with the time scale of heating in ion traps. In order to be able to measure the quantum phase transition, the heating and decoherence time scale of the trap, $T_{\rm decoh}$ should be such that $\delta\nu T_{\rm decoh}\ll 1$, which leads to demanding conditions for the existing trapping setups, see for instance the discussion in Ref.~\onlinecite{Heat-IonTraps,Blatt_Review}.

Another requirement on the experimental setup is the possibility to reduce the temperature $T$ below the energy scale characteristic of the gap $\Delta\sim h$. Using Eq. (\ref{h_UKUP}), the estimated restriction on $T$ (defined for convenience in units of milli-Kelvin) is therefore given by
\begin{equation}
T[{\rm mK}]\ll \frac{10^{3}C_h(U_K^2U_P)^{1/3}}{k_B}\; .
\label{T_UKUP}
\end{equation}
To derive a numerical estimate, we again employ Eqs. (\ref{U_K}), (\ref{U_P}) and rewrite the mass and inter--ion spacing in terms of $n_A$, $a_0$. This yields
\begin{equation}
T[{\rm mK}]\ll 0.25\left(\frac{1}{n_A^2a_0^5}\right)^{1/3}\; ,
\label{T_num}
\end{equation}
implying an upper bound of order $\sim 0.1$mK for protons to
$\mu$K for Magnesium ions. These values are challenging for large crystals, but could be accessed when, for instance, only the transverse motion is cooled (Methods for cooling ion Coulomb crystals to ultralow temperatures are discussed in Ref.~\onlinecite{Eschner03,Wunderlich05}).

We note that the condition on $\delta\nu$ can be relaxed in
the presence of screening with permittivity
$\varepsilon>\varepsilon_0$, with $\varepsilon_0$ the vacuum
permittivity: in this case the potential energy $U_P$ decreases as
$\varepsilon_0/\varepsilon$, and the upper bound on $\delta\nu$
increases with $\varepsilon^{2/3}$. It is interesting to note
that, correspondingly, the condition of the temperature becomes
more restrictive, as the upper bound decreases with
$\varepsilon^{-1/3}$.  This situation could possibly be realized,
for instance, in the case of an ion chain embedded in a crystal,
as in the experiment reported in Ref.~\onlinecite{Drewsen}.

\subsection{Dipolar gases}

Another physical system, where the quantum phase transition can be experimentally observed, are ultracold dipolar gases, such as polar molecules~\cite{Lahaye}. Here, the interaction is repulsive provided the dipoles are aligned perpendicularly to the plane where their motion is studied, for instance, in a very steep trap which freezes out the motion in the perpendicular direction. In Ref.~\onlinecite{Kollath} a zigzag ordering across tubes, containing gases of ultracold polar molecules, has been predicted when the distance between the tubes was below a certain value. Long-range order is found in the presence of an additional external periodic potential which localizes the dipoles in an equidistant array in one dimension and could be realized by means of an optical lattice.
The regime here studied would then correspond to the realization of a Mott-insulator state of the polar molecules in a one dimensional optical lattice,
where the transverse direction must be confined by a dipole trap, whose steepness is changed in order to reach several points of the phase diagram~\cite{Lahaye}. An estimate of the parameters, which allow to access the low dimensional, quasi-crystalline region, has been presented in Ref.~\onlinecite{Astrakharchik}.

\begin{figure}
\centering
{\includegraphics[width=8cm]{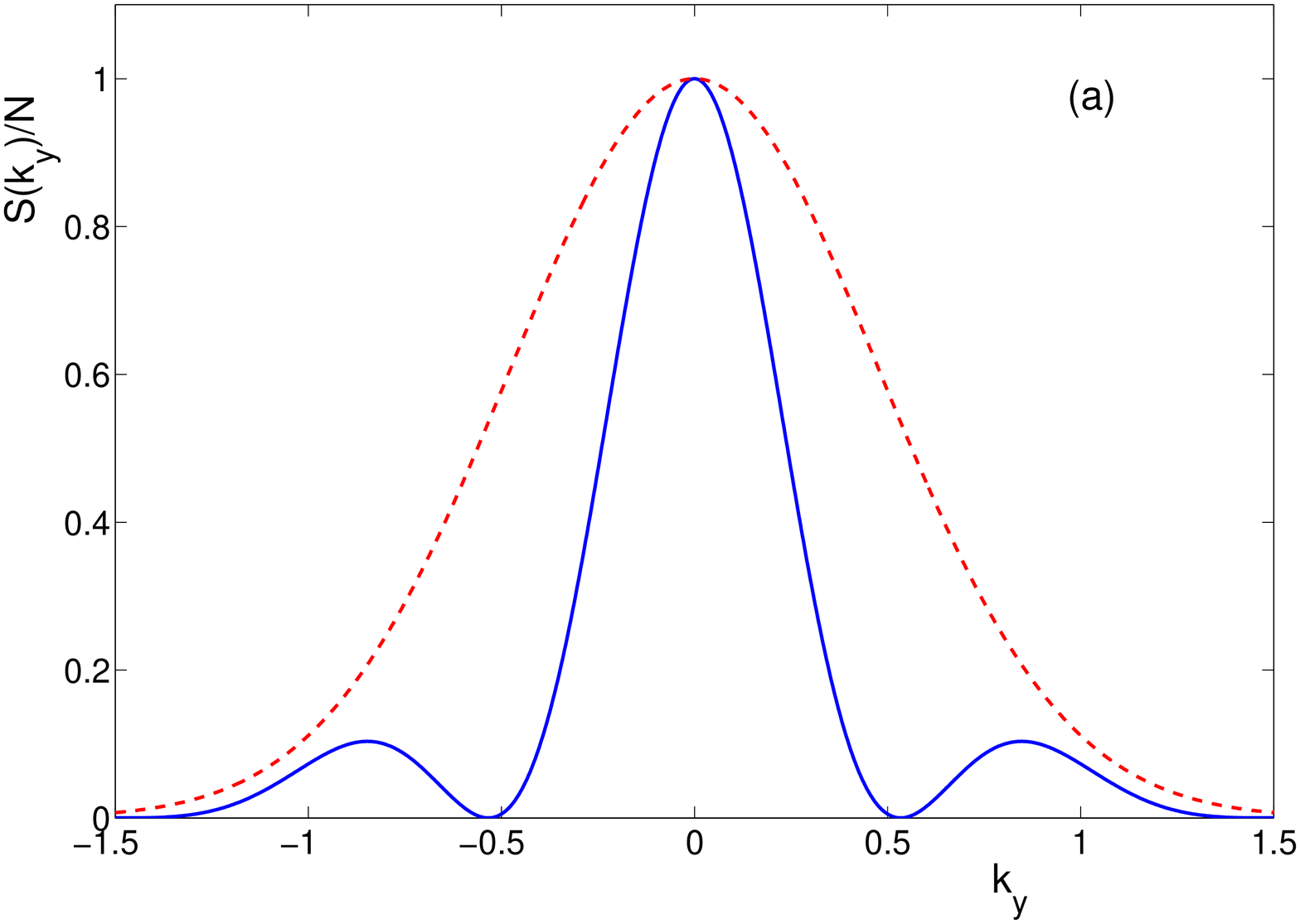}}
{\includegraphics[width=8cm]{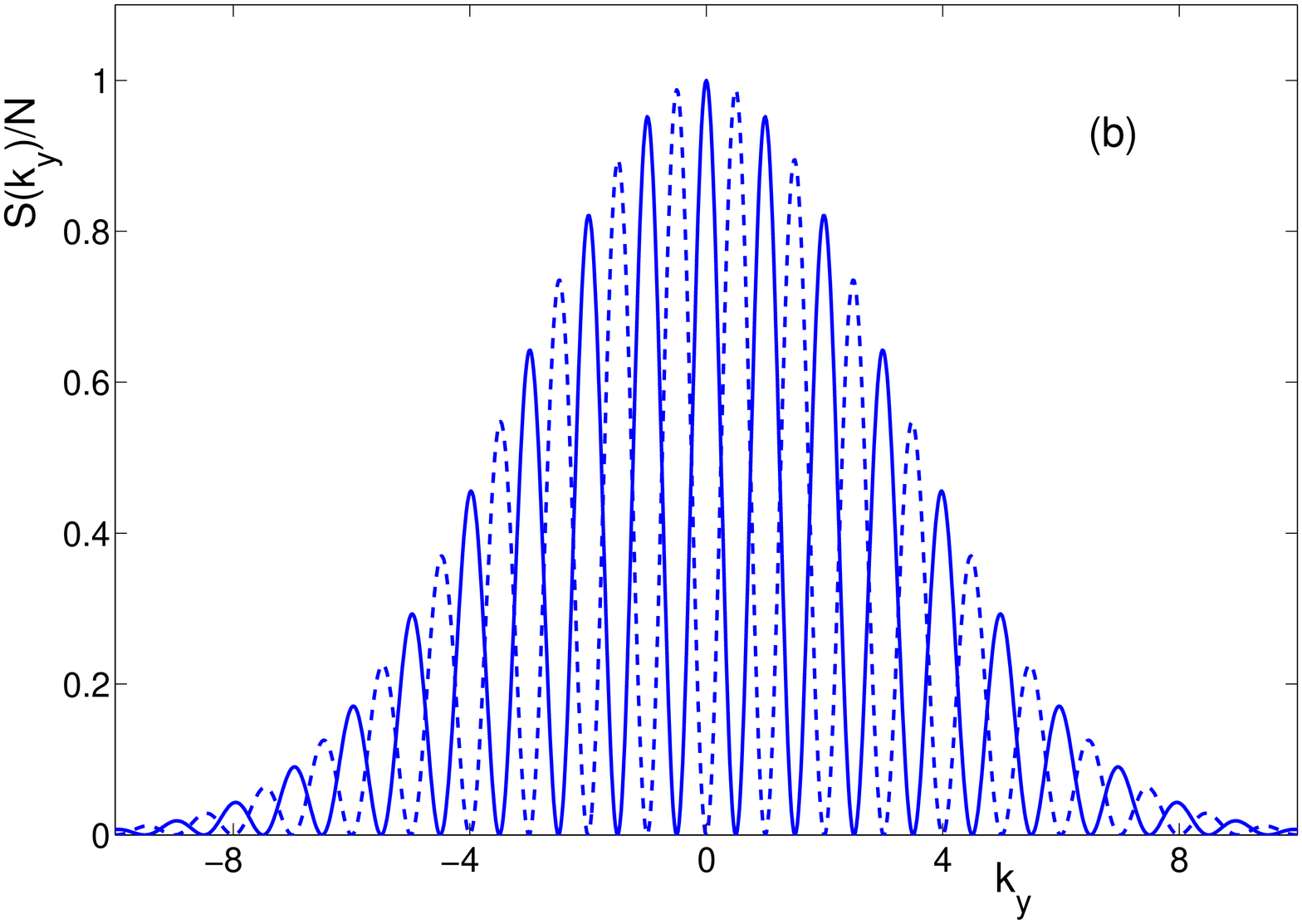}}
\caption[]{(color online) Structure form factor $S(k_y)/N$ as a function of $k_y$
(in units of the transverse length $\phi_0a/\pi$): (a) For an ion string ($\epsilon<\epsilon_c$).
The red  dashed curve corresponds to the classically stable regime (i), where $S(k_y)$ is given
by $S_{\rm lin}$ [Eq. (\ref{S:linear})]; the
blue solid curve corresponds to the ion string in the quantum disordered phase regime (ii)
($S_{\rm dis}$ in Eq.~(\ref{S:disorder}) with $\phi_0/l=1.5$).
(b) For a zigzag structure, regime (iii) (with $\phi_0=10l$). The dashed and solid lines correspond to $S_{\rm zz}(k_y)$ [Eq.
(\ref{S:zigzag})] when $k_x=\pi n/a$ with $n$ even and odd,
respectively.}\label{Fig:2}
\end{figure}

\subsection{Measurement: structure form factor}

In atomic and molecular systems, the measurement of the disordered
phase, where quantum fluctuations dominate in the transverse
direction, can be performed by means of the structure form factor,
here defined by the expression
\begin{equation}
S({\bf k})=\frac{1}{N}\int {\rm d}{\bf x}\int {\rm d}{\bf x'}{\rm
e}^{{\rm i}{\bf k}\cdot({\bf x}-{\bf x'})} \langle n({\bf
x})n({\bf x'})\rangle
\end{equation}
where $n({\bf x})=n(x,y)$ is the density of atoms in the two dimensional plane.
For wave vectors in the limit $k_x\rightarrow k_0$, where $k_0=\pi/a$ is the wave number of the zigzag, $S({\bf k})=S_f+S_0\delta(k_x-k_0)$ in which $S_0$ is proportional to the
squared order parameter, $S_0\propto \phi_0^2$, while $S_f$ is the
contribution of the fluctuations and is proportional to the
isothermal susceptibility $\chi$ of the corresponding Ising
system~\cite{QPTbook,Ma}. Approaching the quantum critical point
for $T\ll\Delta$ (i.e., for $\epsilon\rightarrow \epsilon_c$ below
the dashed lines in Fig. 1), $\chi\sim
|\epsilon_c-\epsilon|^{-\gamma}$ and $\phi_0^2\sim
|\epsilon_c-\epsilon|^{2\beta}$ in the ordered, zigzag phase
($\epsilon> \epsilon_c$), where $\beta=1/8$ and $\gamma=7/4$ are
the exponents of the classical two dimensional Ising model. In the
critical region $T\gg \Delta$ (shaded area in Fig. 1), these
behave as $\chi\sim T^{-7/4}$, $\phi_0^2\sim T^{1/4}$
respectively, where we have used the critical scaling of the
correlation length $\xi\sim 1/\Delta\sim
|\epsilon_c-\epsilon|^{-\nu}$ with $\nu=1$ ~\cite{QPTbook}.
Note that the predicted critical behavior is strictly valid for an
infinite system and assumes a uniform density of particles. However, in
practice it is expected to be approximately valid, e.g. in the
center of a linear Paul trap of ions, when the variation of the density in
a large region is negligible within a correlation length.

The distinct regions in the phase diagram (Fig. 1) can be
identified experimentally also away from criticality (i.e., for
${\bf k}$ that is sufficiently different from ${\bf k_0}$) with
the help of light scattering at the appropriate wave vector. The
structure factor will take a different form in each of three
cases: {\it (i)} the linear (string) chain ($\epsilon<0$), {\it (ii)} the
quantum disordered phase ($0<\epsilon<\epsilon_c$), and {\it (iii)} the
zigzag configuration ($\epsilon >\epsilon_c$) for $N\gg 1$ atoms.

In regime {\it (i)}, where the ion string is assumed, the atomic density
can be written as $n(x,y)=n_x(x)n_y(y)$ such that $n_x(x)$ is the
sum of well spatially-separated Gaussians, centered in $ja$ and
with width $l_x\ll a$, while $n_y(y)$ is a Gaussian with width
$l_y=la$ such that $l\ll 1$. The structure form factor is
then given by
\begin{equation}
\label{S:linear} S_{\rm lin}(k_x,k_y)\simeq N \sum_n
\delta_{k_x,G_n}{\tilde S}(k_x,k_y)
\end{equation}
with
\begin{equation}
{\tilde S}(k_x,k_y)={\rm e}^{-k_x^2 l_x^2/2}{\rm
e}^{-(k_ya)^2l^2/2}
\end{equation}
and corresponds to $\delta$-peaks along the $x$-axis at the
location of the vectors of the reciprocal lattice, $G_n=2\pi n/a$,
where the height of the peaks is embedded in a gaussian function
of width $\sqrt{2}/l_x$, while the dependence on $k_y$ has the
form of a gaussian function of width  $\sqrt{2}/(la)$
(see Fig.~\ref{Fig:2}(a)).

We next consider the zigzag ordered phase regime {\it (iii)}.
Denoting by $y_0=\phi_0a$ the transverse displacement from the
$x$-axis, for $\phi_0\gg l$ we obtain
\begin{equation}
S_{\rm zz}(k_x,k_y)\simeq 4N \sum_n \delta_{k_x,\frac{\pi
n}{a}}{\tilde S}(k_x,k_y)\cos^2\left(\frac{\pi
n}{2}-k_ya\phi_0\right) \label{S:zigzag}
\end{equation}
which provides an interference pattern in the $y$-direction, with
a maximum or a zero at $k_y=0$ depending on whether the
corresponding $x$ component is even or odd, as displayed in
Fig.~\ref{Fig:2}(b). We note that the appearance of an interference pattern for $n$ {\it odd}
(i.e., the doubling of the unit cell) is a primary signature of this ordered phase. In both cases, the pattern is
characterized by peaks separated by $ \pi/a\phi_0$.
These expressions are in agreement with the results found
numerically for the linear and zigzag configuration in ultracold
dipolar gases~\cite{Astra:1}.

Finally, we focus on regime {\it (ii)}, namely the phase where the
zigzag order found in the framework of the classical theory is
destroyed   as a result of quantum fluctuations, and the
disordered linear chain is found.  This regime is characterized by
a wave function whose width in the transverse direction is of the
order of the distance between the two minima of the potential
$V_0$ of Eq. (\ref{V_loc}), see for instance the numerical result
reported in Ref.~\onlinecite{Astrakharchik}. For this purpose, we
assume that the density along $y=a\phi$ is given by
$n_y(y)={\mathcal N}|\Psi_+(\phi)|^2/a$ with $\Psi_+(\phi)$ given
in Eq.~(\ref{pm_wf}) and $\mathcal N=1/(2+2\Gamma)$ giving the right normalization, where
 $\Gamma=\exp(-\phi_0^2/l^2)$ is the overlap
factor [Eq. (\ref{overlap})]. The corresponding structure form
factor reads
\begin{equation}
\label{S:disorder} S_{\rm dis}(k_x,k_y)\simeq N \sum_n
\delta_{k_x,\frac{2\pi n}{a}}{\tilde
S}(k_x,k_y)\left(\frac{\Gamma+\cos(k_ya\phi_0)}{\Gamma+1}\right)^2
\end{equation}
where non-locality is here in the dependence on the parameter
$\Gamma$, giving the overlap between the two wavefunctions in the
two wells. This result was already obtained for the case of a
single atom in a double well potential~\cite{Canizares}, and can
be understood as an interference pattern arising from a photon,
which is elastically scattered by one of the atoms. In the
disordered phase the scattered photon will follow two possible
pathways, associated with the the coherent superposition of the
two positions where the atom can be found with largest
probability. Figure~\ref{Fig:2}(a) displays the
function $S_{\rm dis}$ as a function of $k_y$.
The existence of sidebands (Fig.~\ref{Fig:2}(a)) is the signature of the regime {\it (ii)}.
This demonstrates how one can distinguish the
various regimes and measure the shift of the critical point
resulting from quantum tunnelling.

We also remark that autocorrelation functions of the crystal can be measured by means of Ramsey interferometry, by driving the internal transition of one ion in the chain, as proposed in Ref.~\onlinecite{Chiara}. This method would allow one to extract the critical exponents characterizing the critical behavior. Methods for revealing quantum tunneling between zigzag configurations of few ions have been proposed in Ref.~\onlinecite{Alex}.

\section{Conclusions}
\label{sec:conclude}

In this paper, it has been argued that the structural phase
transition between a chain and a zigzag for ions or atoms
interacting via a power-law interaction falling off as
$1/r^\alpha$ (where $r$ is the distance and $\alpha \ge 1$) can be
mapped on an Ising model in a transverse field. In this way the
classical theory for the phase transition can be extended to
account for quantum fluctuations. Such correspondence was
proposed in the past for electronic systems \cite{MML}. Here the
mapping is explicitly justified deep in the ordered zigzag phase
[see Eq. (\ref{condition})]. Based on symmetry arguments [see Eq.
(\ref{Tloc_gen})] we postulate that this mapping can be extended
to the critical regime. The number of universality classes in two
dimensions is restricted by conformal field theory~\cite{Cardy}.
Therefore, we argue that identifying the symmetry ($Z_2$
in our case) determines the universality class, and
implies that it is the same as of the one dimensional Ising model
in a transverse field. In other words, we assume that
all operators obtained in our theory that differ from the ones of
the conformal field theory of the one dimensional Ising model in a
transverse field are irrelevant in the sense of the
renormalization group. To the best of our knowledge, there are no
existing theoretical tools to verify such statements. However,
this hypothesis is further supported by numerical
studies~\cite{Fisher,Rieger}.

Under these assumptions, we have estimated the
quantum critical point as well as the regime of parameters where
classically zigzag order would be found, but quantum tunneling
and fluctuations suppress it and the linear chain is recovered. In
this calculation it was assumed that the effective potential $V_0$
[Eq. (\ref{V_loc})], obtained in the vicinity of the critical
point according to the classical theory, still holds.

Our analysis allows us to consider possible realizations of a
suitable system in which the predicted quantum phase transition
would be observable. In particular for trapped ion systems, we
identify the physical parameters determining the critical point,
as well as the experimental parameters which are required in order
to access this regime. Schemes for measuring the transition are proposed, which are based on photon scattering.

We remark that this work addresses the two-dimensional case, in which the transition to a zigzag may only happen in a predetermined plane, and give also the full phase diagram including the thermal fluctuations. In three dimensions, where a Goldstone mode is predicted corresponding to rotations of the plane, in which the zigzag is formed around the axis of the chain, one could consider an extension of this mapping. On this basis we conjecture that the system at the quantum phase transition can be mapped to the XY model, and the phase transition is of Kosterlitz-Thouless type.

We finally note that our theory provides a general framework which
allows one to study further effects, such as dynamics at the
quantum critical point. It is worth mentioning that here one
expects the creation of topological defects, i.e. domain walls. It
would be interesting to apply this theory to defect formation,
extending the work done on the classical system
in Ref.~\onlinecite{DelCampo}, and to explore whether these systems allow for
the creation of special kind of entangled states of the ions.

To conclude, using a quantum-field theoretical description we argue that the linear-zigzag instability in two-dimensional systems of trapped ions or polar molecules can be mapped to the one dimensional Ising model in a transverse field. This result demonstrates once more the potentialities offered by these systems as quantum simulators~\cite{Blatt_Review,Lahaye,QMagnet,Monroe:QM}, and more generally for quantum technological applications.

\acknowledgements

We gratefully acknowledge useful discussions with E. Altman, E. Demler, J.
Eschner, R. Fazio, J. Feinberg, M. E. Fisher, Y. Gefen, J. Meyer, M. Raizen and S. Sachdev. This
work has been partially supported by the Israel Science Foundation
(ISF), by the US-Israel Binational Science Foundation (BSF), by
the Minerva Center of Nonlinear Physics of Complex Systems, by the
Shlomo Kaplansky academic chair, by the European Commission (IP
AQUTE, Strep PICC), by the ESF (EUROQUAM, CMMC), and by the
Spanish Ministerio de Ciencia y Innovaci\'on (QOIT,
Consolider-Ingenio 2010; QNLP, FIS2007-66944; Ramon-y-Cajal). E.
S. acknowledges support from the Ministry of Science and
Technology (Grant 3-5792). G. M. acknowledges the German Research
Council (DFG) for support.

\appendix

\section{Semiclassical Energy Eigenvalues for the Double-well Potential}
\label{sec:appA}

In this appendix the eigenvalues of the Hamiltonian $H_0$ [Eq. (\ref{H0_norm})] are calculated
in the semiclassical limit. In this limit the term proportional to
$\phi^2$ in the potential can be ignored. The energy levels can be
found by the Bohr-Sommerfeld rule where the action $I$ is
quantized:
\begin{equation}
\label{action1} I_n=n+const\approx n
\end{equation}
for large $n$. Here, similarly to Eq. (\ref{H0_norm}), we have used units where length scales are normalized by $\lambda =\sqrt{\hbar/m\omega}$ and energy scales by $\hbar\omega/2$. For large energy ($\bar{E}=E/(\hbar\omega/2)$)
where the semiclassical approximation is relevant,
\begin{equation}
\label{action2} I=\oint \bar{p} d{\bar\phi}=\frac{1}{2\pi}\oint
\sqrt{\bar{E}-B{\bar\phi}^4}d{\bar\phi}
\end{equation}
where $\bar{p}$ is the conjugate momentum of the normalized coordinate ${\bar\phi}$, and $B=\hbar g/2m^2\omega^3$ [see Eq. (\ref{B_def})]. The
integral is over a closed trajectory of the particle. By
elementary rescaling one finds:
\begin{equation}
\label{action3}
 I=\left(\frac{\bar{E}}{B^{1/3}}\right)^{3/4}C,~~~~~~~ C\equiv\frac{1}{2\pi}\oint
\sqrt{1-y^4}dy \; .
\end{equation}
Note that the normalized energies scale as $B^{1/3}$, as we found also for low
energies within the approximation $B\gg 1$ [see Eq. (\ref{h_B})]. The semiclassical energies are therefore
\begin{equation}
\label{sen1} \bar{E}_n=B^{1/3}\left(\frac{n}{C}\right)^{4/3}.
\end{equation}
The energy spacings increase with energy. In the physical units
the eigenenergies are given by
\begin{equation}
\label{sen2}
 E_n=\frac{\hbar\omega}{2}\bar{E}_n=\left(\frac{\hbar n}{C}\right)^{4/3}\left(\frac{g}{16m^2}\right)^{1/3}.
\end{equation}

\end{document}